\documentclass[aps,showpacs,twocolumn,showkeys,amsmath,amssymb]{revtex4}
\usepackage{graphicx}
\usepackage{dcolumn}
\usepackage{bm}
\usepackage{mathrsfs}
\usepackage{epsfig}

\newcommand{\N}{N_c}

\begin{document}

\title{Quantum-Number Exotic Tetraquarks at Large $N_c$ in QCD(AS)}

\author{Thomas D. Cohen}
\email{cohen@physics.umd.edu}
\affiliation{Department of Physics, University of Maryland, College Park,
Maryland 20742-4111, USA}

\author{Richard F. Lebed}
\email{richard.lebed@asu.edu}
\affiliation{Department of Physics, Arizona State University, Tempe,
Arizona 85287-1504, USA}

\date{January, 2014}

\begin{abstract}
It is shown that the large $N_c$ limit of QCD with quarks in the
two-index antisymmetric color representation [QCD(AS)] has narrow
tetraquark states with exotic flavor quantum numbers.  They decay into
mesons with a width that is parametrically $O(1/N_c^2)$.  Tetraquarks
with non-exotic quantum numbers mix at leading order with mesons of
the same overall quantum numbers.  QCD(AS) at $N_c=3$ corresponds to
ordinary QCD; its large $N_c$ limit represents an alternative starting
point for a $1/N_c$ expansion to the standard one with quarks in the
fundamental color representation.
\end{abstract}


\maketitle

Our understanding of QCD has been greatly aided by the study of the
large $N_c$ (number of color charges) limit and the $1/N_c$ expansion
introduced by 't~Hooft 40 years ago~\cite{'tHooft:1973jz}.  The limit
takes $N_c \to \infty$ and the coupling constant $g \to 0$ in such a
way as to keep $g^2 N_c$ fixed.  In a few special cases, such as QCD
in 1+1 dimensions~\cite{'tHooft:1974hx} or QCD in the limit of heavy
quark masses~\cite{Witten:1979kh,Cohen:2011cw}, the approach can be
used as a basis for direct quantitative calculations of observables.
However, typically the approach has been more useful in providing a
qualitative understanding of many aspects of hadronic phenomena.

It has generally been thought that exotic hadrons are qualitatively
understood in large $N_c$ QCD, where by ``exotic'' one means hadrons
that do not fit into a classification scheme based upon a simple quark
model.  Quantum-number exotic hadrons are ones that, by quantum
numbers, {\em cannot\/} be $\overline{q} q$ or $qqq$ states.  It has
long been known that glueballs exist as long-lived particles ({\it
  i.e.}, resonances that are parametrically narrow) at large $N_c$ and
that, in this limit, they do not mix with mesons~\cite{Witten:1979kh}.
It has also been known since the late 1990s that quantum-number exotic
``hybrid mesons''---mesons with quantum numbers that cannot be
constructed out a pure $\overline{q} q$ state in a simple quark model
but require a ``valence gluon'', such as $J^{PC}=1^{-+}$---must exist
as long-lived particles~\cite{Cohen:1998jb}.  It has also long been
believed that at large $N_c$ tetraquarks---states composed of two
quarks and two antiquarks---are forbidden at large
$N_c$~\cite{Witten:1979kh,Coleman}.

While the commonly understood situation regarding glueballs and
hybrids remains uncontroversial, recently Weinberg pointed out that
the standard argument against the existence of resonant tetraquark
states is not valid~\cite{Weinberg:2013cfa}.  It is useful to
summarize why tetraquarks were thought to be impossible at large
$N_c$.  Witten and Coleman~\cite{Witten:1979kh,Coleman} both point out
that, when a correlation function for a tetraquark source of the form
$J=\overline{q}q\overline{q}q$ is computed, the leading-order
contribution is $O(N_c^2)$ and consists of two disconnected quark
loops, each one of which has the quantum numbers of an ordinary meson,
and when cut, has a color-singlet $\overline{q} q$ structure.  From
this argument, it was concluded that tetraquark sources produce only
two-meson states and nothing else.  However, as Weinberg observed,
this conclusion does not follow: The leading {\em connected\/}
contribution is $O(N_c^1)$ and is not of a two-meson character;
nothing in this argument excludes a tetraquark pole associated with
it.

A nice way to see that Weinberg's critique is correct is to consider
the case in which the tetraquark source is a vector-isovector of the
form $J^\mu_a({\bf x})= \epsilon_{a b c}\left ( \overline{q}({\bf x})
  \gamma_5 \tau_b q({\bf x}) \right ) \partial^\mu \left (
  \overline{q}({\bf x}) \gamma_5 \tau_c q({\bf x}) \right )$, which
has the quantum numbers of the $\rho$ meson.  Its leading-order
two-point correlation function is indeed represented as a disconnected
$O(N_c^2)$ diagram consistent with two pions in a vector-isovector
configuration.  However, one cannot conclude from this fact that there
is no narrow vector-isovector hadron in the theory.  Indeed, the
$\rho$ meson exists, couples to the source, and contributes to the
correlation function at $O(N_c^1)$.  Similarly, in the case of a
quantum-number exotic tetraquark channel ({\it i.e.}, one whose
quantum numbers cannot be obtained from a pure $\overline{q} q$
state), one cannot conclude just based on the fact that the
disconnected part of the correlator couples to two mesons that no
tetraquark state exists.

Witten~\cite{Witten:1979kh} gives a second argument that tetraquarks
cannot exist as narrow resonances at large $N_c$: Meson-meson interactions are
weak, with scattering amplitudes scaling as $N_c^{-1}$, and hence a
two-meson interacting state does not have the strength to form a bound
or resonant tetraquark.  However,  this argument is also spurious.
Consider again two pions with vector-isovector quantum numbers.
Despite the fact that the interaction is weak at large $N_c$, they do
in fact resonate into a narrow $\rho$ meson.  In a similar manner,
narrow [width $O(N_c^{-1})$] quantum-number exotic tetraquarks coupled
to two mesons with a strength $O(N_c^{-1/2})$ are fully compatible
with meson-meson interactions whose scattering amplitudes scale as
$N_c^{-1}$.

For many observables, the large $N_c$ world is known to behave
similarly to the physical world of $N_c=3$.  Thus, one may be more
likely to interpret some scalar mesons such as the $f_0$(980) as
tetraquarks (as has been commonly suggested over the
years~\cite{Jaffe:1976ig,Close:2002zu,Braaten:2003he,Close:2003sg,
  Maiani:2004uc,Bignamini:2009sk,Ali:2011ug,Achasov:2012kk,
  Friedmann:2009mz}), if tetraquarks exist at large $N_c$.  Weinberg's
analysis has sparked some interesting work on tetraquarks at large
$N_c$.  Two notable results are the observation by Knecht and
Peris~\cite{Knecht:2013yqa} that, if narrow tetraquarks do exist at
large $N_c$, the parametric dependence of the width on $N_c$ depends
upon the flavor content of the state, and Lebed's
demonstration~\cite{Lebed:2013aka} that the existence of narrow
tetraquarks requires a rather subtle $N_c$ dependence of the coupling
of paired bilinear sources to the tetraquark state in the limit in
which the bilinear sources approach the same spatial point.

Weinberg's analysis does not resolve a central question.  It shows
that previous attempts to rule out tetraquarks at large $N_c$ are
flawed, but it does not show that tetraquarks {\em do\/} exist.  The
purpose of the present note is to show that, while the status of
tetraquarks in the most common extrapolation from $N_c=3$ to large
$N_c$ remains unresolved at present, there exists a different but
equally valid extrapolation in which narrow tetraquarks can be shown
necessarily to exist.  The extrapolation in question puts the quarks
into the two-index antisymmetric color
representation~\cite{Corrigan:1979xf,Armoni:2003gp,Armoni:2003fb,
  Armoni:2004uu,Armoni:2005wt} (rather than the color fundamental
representation), and so is often denoted QCD(AS)\@.  At $N_c=3$ the
two-index antisymmetric representation is three-dimensional, and the
theory is identical to QCD\@.  However, the extrapolation to large
$N_c$ is different and forms the basis for a distinct $1/N_c$
expansion.  A principal difference between the two expansions is that
quark loops are not suppressed in QCD(AS), which leads to a different
$N_c$ counting for hadronic vertices involving mesons and to
leading-order glueball-meson mixing.  There has been considerable
interest in QCD(AS) at large $N_c$ due to its beautiful formal
properties, including the emergence of various
dualities~\cite{Armoni:2003gp,
  Armoni:2003fb,Armoni:2004uu,Armoni:2005wt}.  At least for the case
of baryons~\cite{Bolognesi:2006ws,Cherman:2006iy,Cohen:2009wm}, it can
be shown that mass relations based on QCD(AS) have considerable
phenomenological predictive
power~\cite{Cherman:2009fh,Cherman:2012eg}, as do relations based on
the more standard variant.  In general, one expects the expansion that
does the better job describing the data for $N_c=3$ to depend upon the
observable.

Since QCD(AS) includes quark loops at leading order, one expects that
``ordinary'' mesons and tetraquarks might be mixed.  Thus,
distinguishing between mesons and tetraquarks can be problematic.
Here we focus on ``true'' tetraquarks---namely, ones that, at leading
order in the $1/N_c$ expansion, contain only components with at least
two quarks and two antiquarks, and show that such states must exist as
narrow hadrons at large $N_c$ in QCD(AS)\@.  We accomplish this
separation by considering states with so-called exotic quantum
numbers---states that, by construction, cannot be composed of a single
$\overline{q} q$ pair, such as an isospin-two hadron.

\begin{figure}
\begin{center}
\includegraphics[width=3.4in]{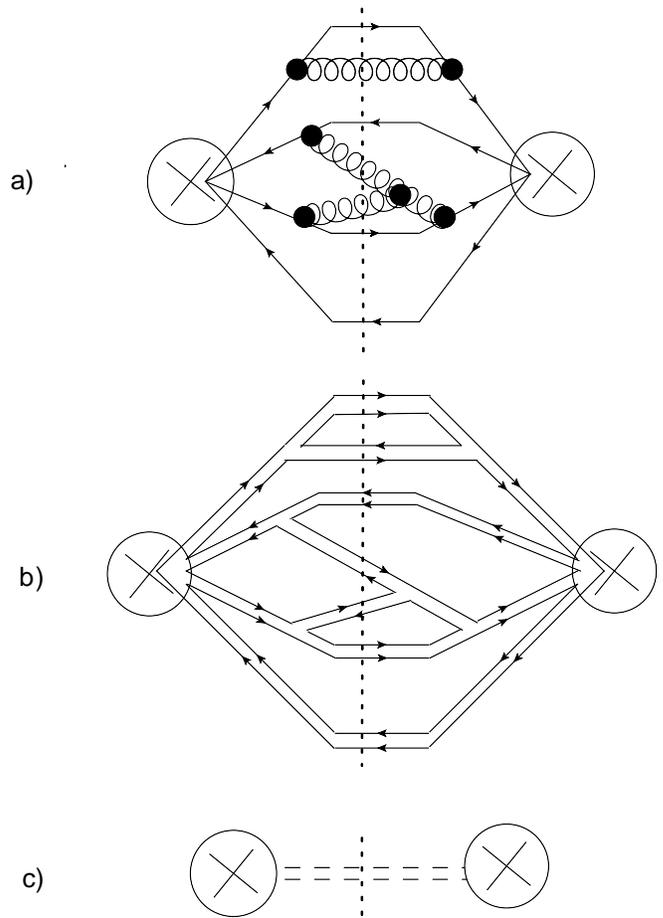}
\caption{Diagram (a) indicates a typical planar Feynman diagram that
  contributes to the leading-order (in $N_c$) two-point correlation
  function for sources of the form of Eq.~(\ref{source}).  The circle
  with a cross indicates the source.  Diagram (b) shows the 't~Hooft
  color-flow diagram associated with diagram (a).  Diagram (c) is a
  hadronic-level depiction, indicating that the leading-order behavior
  is associated with the propagation of single hadron states.  The
  vertical short-dashed lines indicate a cut of the diagram associated
  with one particular intermediate state.}\label{tetra_AS_cut}
\end{center}
\end{figure}
\begin{figure}
\begin{center}
\includegraphics[width=4.5in]{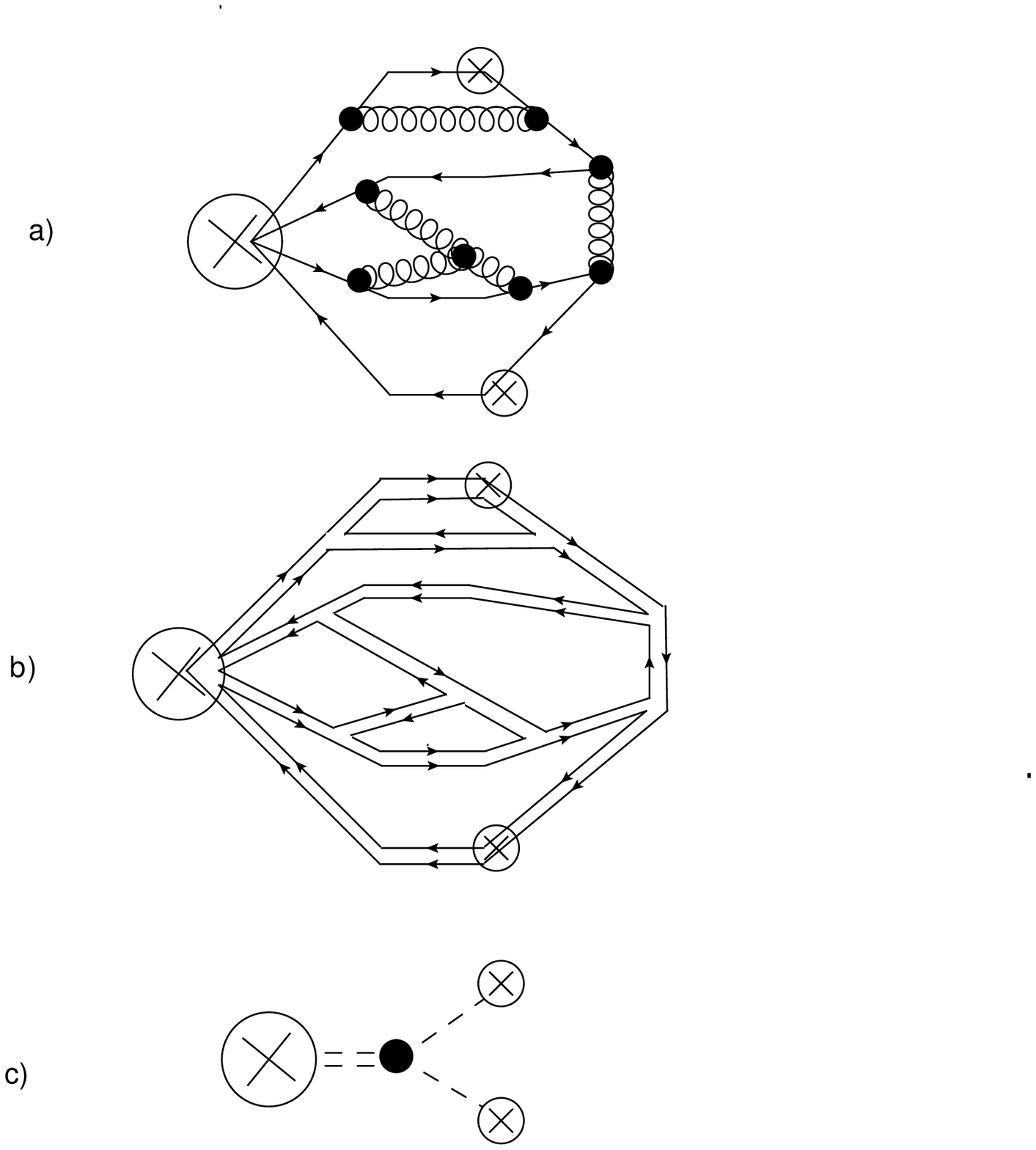}
\caption{Diagram (a) indicates a typical planar Feynman diagram that
  contributes to the leading-order (in $N_c$) three-point correlation
  function for a tetraquark source of the form of Eq.~(\ref{source})
  and two quark bilinear sources.  The large circle with a cross
  indicates the tetraquark source, while the smaller ones indicate the
  bilinear sources.  Diagram (b) shows the 't~Hooft color-flow diagram
  associated with diagram (a).  Diagram (c) is a hadronic-level
  depiction of the leading-order behavior.}\label{tetra_mm_AS}
\end{center}
\end{figure}

The basic strategy is to use the same sort of diagrammatic analysis
commonly used in standard large $N_c$ studies of hadrons and to make
appropriate modifications.  One can repeat the standard double-line
color-flow analysis of 't~Hooft\cite{'tHooft:1973jz}, but now quarks
as well as gluons are represented by double lines.  In contrast to the
adjoint-representation gluons, for the quarks both color lines flow in
the same direction.  Consider the two-point correlation function for a
tetraquark source operator of the following sort:
\begin{equation}
J({\bf x})= \! \sum_{A,B; \; a,b,c,d} \! C_{A , B} \,
\overline{q}^{a b} \, ({\bf x}) \Gamma_A  \, q_{b c} ({\bf x}) \;
\overline{q}^{c d} ({\bf x}) \, \Gamma_B  \,q_{ d a} ({\bf x}) ,
\label{source}
\end{equation}
where lowercase letters represent fundamental color indices, the quark
fields are antisymmetric in color $(q_{a b}=-q_{b a})$, explicit
flavor and Dirac indices for the quarks are suppressed, and
$\Gamma_{A,B}$ represent matrices in Dirac and flavor space.
Spin and flavor quantum numbers are fixed by the choice of $C_{A B}$.
The key point is that the source $J$ as a whole is a color singlet,
but the colors couple the quarks in such a way that one cannot split
$J$ into two color singlets  for $N_c > 3$.  Note that
  color-fundamental quarks cannot be entangled in this way for any
  $N_c$, since Fierz reordering always allows such quarks with
  contracted color indices to be combined into color-singlet bilinears.

The two-point correlation function of the $J$'s is dominated at large
$N_c$ by planar diagrams connecting the two sources.  As the sources
involve four separate color indices that are summed over, one expects
that these diagrams scale at leading order as $N_c^4$, which is indeed
the case.  As an example, consider diagram (a) in
Fig.~\ref{tetra_AS_cut}, which contains 6 coupling constants.  The
color flow is shown in diagram (b), which has 7 color loops.  The 7
color loops yield a factor of $N_c^7$ while the 6 coupling constants
contribute a factor of $N_c^{-3}$, yielding an overall scaling of
$N_c^4$, as advertised.  More generally, any diagram can be
constructed by starting with a skeleton of no gluons and the minimum
number of quark loops, and then adding in planar gluons or planar
quark loops one at a time.  Each of them adds a color loop (a factor
of $N_c$) and two coupling constants (a factor of $N_c^{-1}$), and
hence does not alter the $N_c$ counting.

The central issue is the color structure of the states created by the
source.  Consider diagram (a) of Fig.~\ref{tetra_AS_cut} in more
detail.  The vertical line represents a possible cut that exposes the
intermediate state created.  Its color structure is illustrated in
diagram (b); we see that the color structure of the quarks and gluons
making up that state is $\overline{q}^{a b} A_b^{\; \, c} q_{c d}
A^d_{\; e} A^e_{\, f} \overline{q}^{f g} q_{g a}$.  The key point is
that it is a single-color trace object.  It cannot be split into two
separate color-singlet combinations except due to subleading
contributions in $1/N_c$ in which two colors accidentally coincide.
If one assumes confinement so that all quarks and gluons are bound
into hadrons, this observation means that contributions from this cut
of this diagram correspond to a single hadron.  Moreover, it is easy
to see that this result is generic: All cuts of all leading-order
diagrams using the source $J$ have a single-color trace structure.
Thus, one concludes that the correlation function at leading order is
saturated by single-hadron states; this result indicated by diagram
(c) of Fig.~\ref{tetra_AS_cut}.  If the source creates states that
include ones with exotic quantum numbers, one concludes that at large
$N_c$ quantum-number exotic tetraquarks must exist as narrow hadrons
in the theory.  This is the principal result of this note.

The parametric dependence upon $N_c$ of the interaction of tetraquarks
with themselves and with other hadrons can be determined by studying
higher-point correlation functions.  Note that from the analysis
above, $J$ creates a free [propagator $\sim N_c^0$] tetraquark with an
amplitude $\sim N_c^2$ in QCD(AS), while standard meson and glueball
sources create hadrons with an amplitude of $\sim N_c^1$.  Consider,
as a concrete example, the tetraquark-meson-meson vertex.  One might
think that a typical diagram contributing to three-point function can
be obtained from a typical contribution to the tetraquark two-point
function by simply removing a tetraquark source and adding two meson
sources.  However, this cannot be done: The tetraquark source $J$
scrambles the colors of the various sources.  To reconnect the colors
when removing $J$, one needs to add at least one gluon exchange, as in
going from the diagrams in Fig.~\ref{tetra_AS_cut} to those in
Fig.~\ref{tetra_mm_AS}.  Note that, in adding the gluon, one does not
change the number of color loops as compared to the two-point function
[there are still 7 in diagram (b) of Fig.~\ref{tetra_mm_AS}], but the
graph has two additional coupling constants [there are 8 in diagram
(a) of Fig.~\ref{tetra_mm_AS}], which costs an additional factor of
$N_c^{-1}$.  Thus, the overall $N_c$ scaling of the diagram is
$N_c^3$.  This scaling is generic; the leading contribution to
three-point functions with one tetraquark source and two ordinary
meson sources is $N_c^3$.

At the hadronic level, this correlation function is dominated by a
single tetraquark created by the source $J$ with amplitude $\sim
N_c^2$ and each meson source producing a single meson with amplitude
$\sim N_c$, for a total of $N_c^4$.  The hadrons propagate and
interact at a vertex, as in diagram (c) of Fig.~\ref{tetra_mm_AS}.
Together, the amplitudes for creating the hadrons ($\sim N_c^4$)
folded in with the propagation of each hadron ($\sim N_c^0$) and the
tetraquark-meson-meson interaction vertex must yield the full
correlation function ($\sim N_c^3$).  Thus, the tetraquark-meson-meson
interaction vertex must scale as $1/N_c$, and the decay width of a
tetraquark into two mesons scales as $1/N_c^2$---which turns out to be
the leading behavior for the tetraquark width: As noted earlier, at
large $N_c$ the tetraquark becomes stable.  Using similar reasoning,
it is easy to show that $\Gamma_n$, a general hadronic vertex with $n$
hadrons (tetraquarks, glueballs, hybrids, and mesons) scales with
$N_c$ as
\begin{equation}
\Gamma_n \sim N_c^{2- n}  \; .
\label{scaling}
\end{equation}
In deriving Eq.~(\ref{scaling}), the key first step is to show that
the $N_c$ scaling of a diagram $D$ containing $n_T$ tetraquark sources
and any number of meson, hybrid, and gluon sources scales as
\begin{equation}
D \sim N_c^{2 + n_T}. 
\end{equation}
One consequence of Eq.~(\ref{scaling}) is that glueballs, hybrids and
mesons all mix at leading order in QCD(AS), if allowed by quantum
numbers (as can be seen from the two-point functions).  Since the
tetraquarks are exotic  they do not mix with other hadrons.  Note that  
tetraquarks sourced by a variant of Eq.~(\ref{source}) in which the bilinears 
are separately color singlets may still have exotic quantum numbers, but 
they would mix with conventional two-meson states at leading order, 
as discussed above.

The analysis goes through without substantial formal changes for the
case of non-exotic quantum numbers, and Eq.~(\ref{scaling}) continues
to hold.  However, tetraquarks and mesons with non-exotic overall
quantum numbers can mix at leading order, as indicated by
Eq.~(\ref{scaling}).  This is hardly surprising: Quark loops are not
suppressed in QCD(AS)\@.  A $\overline{q} q$ pair of the same flavor
in a tetraquark can annihilate into a gluon, leaving behind a single
$\overline{q} q$ pair.  In the case of non-exotic quantum numbers,
such a pair necessarily occurs.  Thus, it is generally not possible to
distinguish between tetraquarks and mesons with non-exotic quantum
numbers.

One might hope that, if the theory has an exact flavor symmetry, there
exist ``true'' tetraquarks with non-exotic overall quantum numbers
that (at leading order) only contain components with two or more
$\overline{q} q$ pairs and thus do not mix at leading order with
ordinary mesons.  This scenario requires that the annihilation
amplitude somehow cancels due to symmetry.  However, such
configurations do not appear to exist.  Such a tetraquark would be
natural for a source in which the color is in the configuration of
Eq.~(\ref{source}), while the flavor is in a configuration such that
neither $\overline{q} q$ pair has a flavor-singlet component.  At
first sight, constructing such a state seems easy: For the case of two
degenerate flavors, simply put each pair into an isovector
configuration and then combine them to total isospin zero or one,
yielding non-exotic overall quantum numbers.  However, there are two
distinct ways to form $\overline{q} q$ pairs since each quark could
pair with either antiquark; to prevent annihilation of a pair, the
flavor configuration must be such that, with either pairing, no flavor
singlet component exists for either pair.  But no flavor configuration
with this property exists.

This discussion suggests that the question of whether the $f_0$(980)
or other mesons are tetraquarks is not entirely well posed at large
$N_c$ in QCD(AS); such states are mixed with both tetraquark and
ordinary mesonic components, and the mixing is not parametrically
suppressed.  However, a scenario in which the mixing is {\em
  numerically\/} small and the state is dominated by the tetraquark
configuration is fully consistent with what is known about QCD(AS) at
large $N_c$.  Finally, it should be noted that this type of analysis
does not only imply tetraquarks.  Hexaquarks, octaquarks and higher
configurations with exotic quantum numbers exist as narrow states in
QCD(AS) at large $N_c$, while such states with non-exotic quantum
numbers mix with ordinary hadrons.

\begin{acknowledgments}
The authors gratefully acknowledge support by the U.S.\ Department of
Energy under Grant DE-FG02-93ER-40762 (TDC) and by the National
Science Foundation under Grant PHY-1068286 (RFL).
\end{acknowledgments}

\end{document}